\documentclass[12pt]{article}
\usepackage{epsfig}

\def\appendix#1{
  \addtocounter{section}{1}
 \setcounter{equation}{0}
  \renewcommand{\thesection}{\Alph{section}}
 \section*{Appendix \thesection\protect\indent \parbox[t]{11.715cm} {#1}}
  \addcontentsline{toc}{section}{Appendix \thesection\ \ \ #1}
  }

\newcommand{\newsection}{
\setcounter{equation}{0}
\section}

\newcommand{\eq}[1]{\begin{equation} #1 \end{equation}}
\newcommand{\ar}[1]{\begin{eqnarray} #1 \end{eqnarray}}
\newcommand{\tr}{\mathop{\mathrm{tr}}\nolimits}

\def\e{{\,\rm e}\,}

\def\D{\delta}

\newcommand{\br}[1]{\left( #1 \right)}
\newcommand{\vev}[1]{\left\langle #1 \right\rangle}
\newcommand{\rf}[1]{(\ref{#1})}
\newcommand{\non}{\nonumber \\*}
\def\N{${\cal N}=4$ }

\def\t{\sqrt{\lambda}}
\def\q{\theta}
\def\o{{\cal O}}

\def\nco{\frac{\vev{W(C)\,\o^I_k}}{\vev{W(C)}}}
\def\co{\vev{W(C)\,\o^I_k}}
\def\c{{\cal C}}

\def\l0{~~~(\lambda\rightarrow 0)}
\def\linf{~~~(\lambda\rightarrow \infty)}
\def\s{\varphi}

\title{\hfill{\small ITEP-TH-28/01}\\~\\
More exact predictions of SUSYM for string theory}

\author{Gordon W. Semenoff$^{~1}$ and K. Zarembo$^{~1,2,3}$
\\~~\\
1. Department of Physics and Astronomy, \\
2. Pacific Institute for the Mathematical Sciences\\
University of British Columbia, \\
Vancouver, British Columbia, Canada V6T 1Z1.
\\~~\\
3. Institute of Theoretical and Experimental Physics,\\
B. Cheremushkinskaya 25, 117259\\ Moscow, Russia}

\begin{document}           

\maketitle                 

\abstract{We compute the coefficients of an infinite family of 
chiral primary operators in the local operator expansion of a circular
Wilson loop in ${\cal N}=4$ supersymmetric Yang-Mills theory.  The
computation sums all planar rainbow Feynman graphs.  We argue that
radiative corrections from planar graphs with internal vertices cancel
in leading orders and we conjecture that they cancel to all orders in
perturbation theory.  The coefficients are non-trivial functions of
the 'tHooft coupling and their strong coupling limits are in exact
agreement with those previously computed using the AdS/CFT
correspondence. They predict the sub-leading orders in strong coupling
and could in principle be compared with string theory calculations.}
\newpage

\setcounter{page}{2}
\newsection{Introduction and summary of main results}

The idea that a quantized gauge theory could have a dual description
as a string theory has a long history. Recently one concrete
realization of such a duality has emerged. It has been conjectured
\cite{Maldacena:1998re} that there is an exact mapping between \N
supersymmetric Yang-Mills theory (SYM) with gauge group SU(N) on four
dimensional spacetime and IIB superstring theory on background
$AdS_5\times S_5$ with N units of RR flux.

This mapping is most useful in the low energy, weakly coupled limit of
the string theory.  This coincides with the large N 'tHooft limit
\cite{'tHooft:1974jz} of SYM theory, where N is taken to infinity
holding the combination of Yang-Mills coupling constant and N, defined
by $\lambda=g^2$N, fixed and then taking the large $\lambda$ limit.
This projects onto the strong coupling limit of the sum of planar
Feynman diagrams.  In string theory, this coincides with the classical
low energy limit where the string theory is accurately described by
type IIB supergravity on the background space $AdS_5\times S_5$. Some
explicit computations can be done there.  The results can then be
interpreted in terms of the gauge theory using a well-defined
prescription ~\cite{Gubser:1998bc},\cite{Witten:1998qj}.

Though it has been used for many computations of the strong coupling
limits of gauge theory quantities (see
refs.~\cite{Petersen:1999zh}-\cite{Freedman:2000fy} for reviews), it
is difficult to obtain a direct check of the Maldacena conjecture.
The reason for this is the fact that the correspondence with
supergravity computes gauge theory in the large $\lambda$ limit, with
corrections from tree level string effects being suppressed by powers
of $1/\sqrt{\lambda}$ and sometimes computable to the next order.  On
the other hand, the only other analytical tool which can be used
systematically in the gauge theory is perturbation theory which is an
asymptotic expansion in small $\lambda$.  Generally, the only
quantities for which these expansions have an overlapping range of
validity is for quantities which are so protected by supersymmetry
that they do not depend on the coupling constant.

There is, however, one known example of a quantity which is a
non-trivial function of the coupling constant and whose large N limit
is computable and is thought to be known to all orders in perturbation
theory in planar diagrams. That quantity is the circular Wilson loop.
Its expectation value was computed in ref. \cite{Erickson:2000af}.
The contribution of a subset of all Feynman graphs, the planar rainbow
diagrams, were found at each order in $\lambda$ and the sum of all
orders was taken to obtain the result

\begin{equation}
\langle W[{\rm circle}] \rangle
= \frac{2}{\sqrt{\lambda}}I_1(\sqrt{\lambda})
~~
\approx
~~\sqrt{\frac{2}{\pi}}
\frac{e^{\sqrt{\lambda}}}{\lambda^{3/4} } ~{\rm as}~\lambda\to\infty
\label{circle}
\end{equation}
where $I_1(x)$ is a modified Bessel function.  It was also shown
explicitly that the leading corrections to the sum of rainbow diagrams
cancels identically.  It was conjectured that this cancellation would
also occur at higher orders and the result (\ref{circle}) was thus the
exact sum of all planar diagrams.  Some support for this conjecture
was developed in ref.~\cite{Drukker:2000rr}.  They also observed that
the sum over Feynman diagrams could be obtained for all orders in the
$1/N$ expansion and had a beautiful argument that, in the large
$\lambda$ limit, these higher orders produced the expected higher
genus string corrections.

The large $\lambda$ limit in (\ref{circle}) agrees with the
expectation value of the circular Wilson loop which was computed using
the AdS/CFT correspondence in
refs. \cite{Berenstein:1999ij},\cite{Drukker:1999zq} .  If
(\ref{circle}) is indeed an exact result, this provides a non-trivial
check on the validity of the AdS/CFT correspondence.  It gives the
further interesting possibility of comparing corrections at
sub-leading orders in $1/\sqrt{\lambda}$ with string theory
computations.  Investigations of the relevant string theory technique
appears in refs.~\cite{Greensite:1999jw}-\cite{Drukker:2000ep} but
explicit calculations of the $1/\sqrt{\lambda}$ corrections have not
yet been done.

In this Paper we shall report the computation of a series of expansion
coefficients which are related to the circular Wilson loop and chiral
primary operators in \N SYM theory.  The problem that we pose is the
following.  When probed from a distance much larger than the size of
the loop, the Wilson loop operator can be expanded in a series of
local operators with some coefficients
\cite{Shifman:1980ui},\cite{Berenstein:1999ij}:
\begin{equation} 
W[{C}]= \langle W[C]\rangle\sum_{\Delta}
{\cal C}_A R^{\Delta_A} {\cal O}^A(0) 
\label{expansion}
\end{equation} 
where ${\cal O}^A(0)$ is a local operator evaluated at the center of
the loop, $\Delta_A$ is the conformal dimension of ${\cal O}^A(x)$ and
$R$ is the radius of the loop.  The problem is to compute the
coefficients ${\cal C}_A$ in this operator product expansion (OPE).

We shall concentrate on computing the coefficients for a particular
class of chiral primary operators (CPO). We will be able to compute
the contribution of the sum of all planar rainbow graphs to the
coefficients ${\cal C}_A$ in that case.  We are also able to show that
the leading order corrections to this sum, which come from diagrams
with internal vertices, cancels identically.  This leads us to
conjecture that the radiative corrections cancel to all orders and the
sum of planar rainbow graphs gives the exact result.

We find that the coefficients that we compute are non-trivial
functions of the coupling constant.  In the limit of large $\lambda$
they coincide with results of the AdS/CFT correspondence
~\cite{Berenstein:1999ij}.  This gives a large array of non-trivial
functions of the coupling constant which could be compared with string
computations of the strong coupling limit.  There are various reasons
why these computations could be simpler than the $1/\sqrt{\lambda}$
corrections to the expectation value of the Wilson loop itself.
 
The Wilson loop operator in \N SYM theory that is readily computed
using the AdS/CFT correspondence and which has the right
transformation properties under supersymmetry
~\cite{Maldacena:1998im},\cite{Drukker:1999zq} contains the scalar
fields inside the path-ordered exponential:
\eq{\label{php}
W[C]=\frac{1}{N}\,
\tr{\rm P}\exp\left[\oint_{C} d\tau\,\br{i A_\mu(x)\dot{x}_\mu 
+\Phi^i(x)\theta_i|\dot{x}|}\right],
}
where $x_\mu(\tau)$ parameterize the contour ${C}$ and $\q_i$
are Cartesian coordinates of a point on $S^5$: $\q^2=1$.
If the size of the contour ${C}$ is small, the Wilson loop 
can be expanded in local operators as in (\ref{expansion}).

For  primary operators, one can choose a basis where
\begin{equation}
\langle {\cal O}^A(x) {\cal O}^B(y) \rangle = \frac{ \delta^{AB} }{
\left| x-y \right|^{\Delta_A + \Delta_B} }
\label{normalization}
\end{equation}
Then their OPE coefficients can be extracted from the large distance
behavior of connected two-point correlation functions,
\eq{\label{ope}
\frac{\vev{W(C)\,\o^A(L)}_c}{\vev{W(C)}}=\c_A\,
\frac{R^{\Delta_A}}{L^{2\Delta_A}}+\cdots\,
} where $L\gg R$ and the omitted terms are of higher order in
$R^2/L^2$.

The coefficient corresponding to the CPO of lowest conformal
dimension, which in this case is $\Delta=2$, is important as it
determines the correlator of two Wilson loops with large separation
\begin{equation}
\frac{  \left< W[C_1]W[C_2] \right>_c }{
\left<W[C_1]\right>\left<W[C_2]\right> }=
{\cal C}_2^2\left( \frac{R}{L}\right)^4+\ldots
\end{equation}

The coefficients of various CPOs in the expansion of the circular
Wilson loop were calculated in ref.  ~\cite{Berenstein:1999ij} both
perturbatively at $\lambda\sim0$ and at strong coupling,
$\lambda\sim\infty$, using the AdS/CFT correspondence.  Evaluation of
the correlators that define coefficients in the strong-coupling regime
involves a hybrid of the supergravity and the string calculations.

In AdS/CFT, the Wilson loop operator (\ref{php}) naturally couples to
stringy degrees of freedom.  It creates a classical string world-sheet
which is embedded in $AdS_5\times S_5$ and whose boundary is
the contour of the Wilson loop~\cite{Maldacena:1998im},\cite{Rey:1998ik}.

On the other hand, a local operator ${\cal O}^A(x)$ emits one of the
supergravity fields at point $x$.  When it contributes to a correlator
of $O_A(x)$ with the Wilson loop, this supergravity mode propagates
on the background $AdS_5\times S_5$ and is then absorbed by a vertex
operator which must be integrated over the string world-sheet.

\subsection{Dimension two operators}

Let us begin by considering the CPO with smallest conformal dimension,
$\Delta=2$.  It is the symmetric traceless part of a gauge invariant
product of scalar fields,

\eq{\label{dim2}
\o^{ij}=\frac{8\pi}{\sqrt{2}\,\lambda }\tr\br{\Phi^i\Phi^j
-\frac{1}{6}\,\D^{ij}\Phi^2}. } 

This operator is the lowest weight component of a short multiplet of
\N super-conformal algebra. Such chiral primary operators have very special
properties.  The super-conformal algebra guarantees that their
conformal dimensions do not receive radiative corrections, so in this
case the conformal dimension is exactly two.  Furthermore, it is known
that their two and three-point correlation functions are given by the
free field values, that is, that they are independent of the coupling
constant, $g$.  It is known that their four-point functions are
non-trivial, so they are not free fields in disguise
\cite{Lee:1998bx}-\cite{Bianchi:1999ge}.

In (\ref{dim2}), the overall coefficient is chosen to give a canonical
normalization of the two-point function:
\eq{
\vev{\o^{ij}(x)\o^{kl}(y)}=\frac{1}{2}\br{\D^{ik}\D^{jl}+\D^{il}\D^{jk}
-\frac{1}{3}\,\D^{ij}\D^{kl}}\,\frac{1}{|x-y|^4}\,.  } 

The small coupling limit of the correlator of ${\cal O}^{ij}$ with the
Wilson loop is straightforward to obtain.  To leading order in
perturbation theory:

\eq{\label{wcd2}
\frac{\vev{W(C)\,\o^{ij}}}{\vev{W(C)}}=\frac{1}{N}\,
\frac{1}{2\sqrt{2}}\,\lambda\br{\q^i\q^j-\frac{1}{6}\D^{ij}}
\,\frac{R^2}{L^4}\l0.
} The linear dependence on $\lambda$ is an obvious consequence of the
fact that the correlator contains two propagators and one power of
$\lambda$ is cancelled by the normalization.

The AdS dual of the dimension two operator is the negative mass
scalar which is a linear combination of the trace of the metric and
the Ramond-Ramond four-form field.  Its contribution to the OPE of the
circular Wilson loop was calculated in ~\cite{Berenstein:1999ij}:
\eq{\label{scd2}
\frac{\vev{W(C)\,\o^{ij}}}{\vev{W(C)}}=\frac{1}{N}\,
\sqrt{2\lambda}\,\br{\q^i\q^j-\frac{1}{6}\D^{ij}}
\,\frac{R^2}{L^4}\linf .
} Comparing OPE coefficients at strong and at weak coupling, we see
that the scaling with $\lambda$ is different.  The OPE coefficients
are clearly renormalized by radiative corrections. We shall conjecture
that, in the large N limit, this renormalization is entirely due to planar 
rainbow diagrams.  
We shall also obtain the sum of planar rainbows as
\eq{\label{exd2}
\frac{\vev{W(C)\,\o^{ij}}}{\vev{W(C)}}=\frac{1}{N}\,
\sqrt{2\lambda}\,\,\frac{I_2\br{\t}}{I_1\br{\t}}\,
\br{\q^i\q^j-\frac{1}{6}\D^{ij}}
\,\frac{R^2}{L^4}\,,
}
where $I_2$ and $I_1$ are modified Bessel functions. By construction, this
expression reduces to \rf{wcd2} at small $\lambda$.
Since
\eq{\label{asbess}
\lim_{\lambda\rightarrow\infty}\frac{I_k\br{\t}}{I_1\br{\t}}=1
} 
for any $k$, the AdS/CFT prediction \rf{scd2} is also exactly reproduced 
at large $\lambda$. The sum of rainbow diagrams thus interpolates between
perturbative and strong coupling limits of the OPE coefficient.

\subsection{Chiral primary operators}

The $\o^{ij}$ is the first in an infinite sequence of CPOs.  The
operator of dimension $k$ in this sequence is a symmetrized trace of
$k$ scalar fields:
\eq{\label{cpo}
\o^I_k=\frac{(8\pi^2)^{k/2}}{\sqrt{k}\lambda^{k/2}}\,
C^I_{i_1\ldots i_k}\tr \Phi^{i_1}\ldots\Phi^{i_k},
}
where $C^I_{i_1\ldots i_k}$ are totally symmetric traceless
tensors normalized as
\eq{
C^I_{i_1\ldots i_k}C^J_{i_1\ldots i_k}=\D^{IJ}.  } 

The AdS duals of CPOs are Kaluza Klein modes of the $AdS_5$ tachyonic
scalar on $S^5$ and each CPO is associated with a spherical harmonic:
\eq{
Y^I(\q)=C^I_{i_1\ldots i_k}\q^{i_1}\ldots\q^{i_k}.  } Here, we are
following the notation of
refs.~\cite{Lee:1998bx},\cite{Berenstein:1999ij}.

The OPE coefficients  depend on how the operators are normalized.
When comparing perturbative calculations with the AdS/CFT predictions,
we need to use the same normalization.  For operators in short
multiplets of \N supersymmetry, this is easy to achieve, since the two
point correlation functions of such operators do not receive radiative
corrections and can be used to fix normalization.  The coefficient in
\rf{cpo} is chosen to unit normalize the two point function:
\eq{
\vev{\o^I_k(x)\o^J_k(y)}=\frac{\D^{IJ}}{|x-y|^{2k}}\,.
} The same conventions were used in the supergravity calculations of
ref.~\cite{Berenstein:1999ij}.

At weak coupling, the OPE coefficient of the circular Wilson loop
is proportional to $\lambda^{k/2}$, where $k$ is the dimension of
the CPO:
\eq{\label{wc}
\nco=\frac{1}{N}\,
2^{-k/2}\,\frac{\sqrt{k}}{k!}\,\lambda^{k/2}\,\frac{R^k}{L^{2k}}\,Y^I(\q)
\l0.
}
It turns out that AdS/CFT correspondence predicts 
a universal scaling of
the OPE coefficients with $\lambda$
at strong coupling: all of them are proportional to $\t$ independently
of $k$. This can be easily understood by
considering a pair correlator of the Wilson
loops, which is quadratic in OPE coefficients. 
The Wilson loop correlator is 
described by an annulus string amplitude
and therefore 
is proportional to the string coupling $g_s$.  
According to the AdS/CFT dictionary,
$$
g_s=\frac{g^2}{4\pi}=\frac{\lambda}{4\pi N}.
$$
Hence,
OPE coefficients must scale as $\t$. An explicit calculation gives
\cite{Berenstein:1999ij}:
\eq{\label{sc}
\nco=\frac{1}{N}\,
2^{k/2-1}\sqrt{k\lambda}\,\,\frac{R^k}{L^{2k}}\,Y^I(\q)
\linf.
}

Our main result is an expression for correlators of the circular
Wilson loop with CPOs:
\eq{\label{ex}
\nco=\frac{1}{N}\,
2^{k/2-1}\sqrt{k\lambda}\,\,\frac{I_k\br{\t}}{I_1\br{\t}}\,
\frac{R^k}{L^{2k}}\,Y^I(\q)
} which we expect is exact in the large N limit.  Its expansion in
$\lambda$ reproduces \rf{wc}. The strong-coupling limit exactly
coincides with the AdS/CFT prediction (using eq.~\rf{asbess}).

\newsection{Re-summation of rainbow diagrams}\label{rrd}

The Euclidean action of \N supersymmetric Yang-Mills theory is
\begin{eqnarray}
S=\int d^4x \frac{N}{\lambda}{\rm Tr}\left\{ \frac{1}{2}F_{\mu\nu}^2
+(D_\mu\Phi^i)^2-\sum_{i<j}[\Phi^i,\Phi^j]^2 +
\right. \nonumber \\ \left.
+\psi^T\Gamma^\mu D_\mu\psi-i\psi^T\Gamma^i[\Phi^i,\psi] \right\}
\end{eqnarray}
where $(\Gamma^\mu,\Gamma^i)$ are ten dimensional Dirac matrices in
the Majorana-Weyl representation.

We will work in the Feynman gauge where the gauge field propagator
has the form
\begin{eqnarray}
\left< A^{ab}_\mu(x) A^{cd}_\nu(y) \right>_0 = \lambda
\frac{\delta_{\mu\nu} }{8\pi^2
(x-y)^2}\frac{ \delta^{ad}\delta^{bc}}{N}
\end{eqnarray}

\begin{figure}[h]
\hspace*{4cm}
\epsfxsize=8cm
\epsfbox{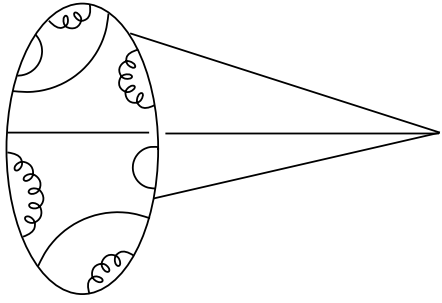}
\caption[x]{\small A typical diagram that contributes to the correlator
of the circular Wilson loop with the dimension $k$ CPO.}
\label{diagr}
\end{figure}

Our calculation of the OPE coefficients begins with summing all planar
rainbow diagrams of the kind shown in fig. \ref{diagr}.  
They contain $k$ scalar propagators connecting the point $L$ to the
Wilson loop and propagators of scalars and gauge fields 
connecting points in segments of the loop.

If $L$ lies on the axis of symmetry of the circle, the scalar
propagators are constants equal to 

$$
\frac{1}{8\pi^2(L^2+R^2)}.
$$
If the origin is displaced from the axis of symmetry, the propagators 
will depend on positions of their endpoints on the circle.
In any case,
we will be interested in the large-distance asymptotics, and the 
propagators can be set to $1/8\pi^2 L^2$, up to corrections of
higher order in $1/L$.

The problem thus reduces to re-summation of rainbow diagrams for each
of the segments of the circle. This problem was solved in
ref.~\cite{Zarembo:2001jp}.  In the following we will review the
salient points involved in finding the solution.  We start with the
dimension two operator \rf{dim2}, when there are only two segments.

\subsection{Dimension two operators}

In this case we have:
\eq{\label{ex1dim2}
\vev{W(C)\,\o^{ij}}= \frac{1}{N}\,         
\frac{8\pi}{\sqrt{2}\,\lambda}
\,\frac{\lambda^2}{(8\pi^2L^2)^2}\,
\br{\q^i\q^j-\frac{1}{6}\D^{ij}}
R^2\, 2\pi\int_0^{2\pi} d\s\, W(\s)W(2\pi-\s), } where $W(\s'-\s)$
denotes the sum of rainbow graphs for a segment of the circle between
polar angles $\s$ and $\s'$.  To compute this sum, we notice that the
sum of the scalar and the gluon propagators between any two points on
a circle does not depend on the positions of these points:
\ar{\label{prp}
\vev{
\br{iA_\mu(x)\dot{x}_\mu
+\Phi^i(x)\theta^i|\dot{x}|}_{ab}
\br{i A_\mu(y)\dot{y}_\mu
+\Phi^i(y)\theta^i|\dot{y}|}_{cd}
}_0&=&\frac{\lambda}{N}\,\D_{ad}\D_{bc}\,
\frac{|\dot{x}| |\dot{y}|-\dot{x}\cdot\dot{y}}{8\pi^2|x-y|^2}
\non
&=&\frac{\lambda}{16\pi^2N}\,\D_{ad}\D_{bc}\,.
}
This observation allows us to replace the field-theory Wick contraction by
the matrix-model average defined by the partition function
\eq{
Z=\int dM\,\exp\br{-\frac{8\pi^2}{\lambda}\,N\tr M^2}.
}
Upon the replacement of $iA_\mu(x)\dot{x}_\mu
+\Phi^i(x)\theta^i|\dot{x}|$ by $M$,
the sum of rainbow diagrams in the segment of the length $\varphi$
reduces to the matrix-model counterpart
of the Wilson loop:
\eq{\label{mm}
W(\s)=\vev{\frac{1}{N}\,
\tr\e^{\s M}}.
}
The matrix model can be viewed as a combinatorial tool which simply
counts the number of planar graphs \cite{Erickson:2000af}.

The Wilson loop in the matrix model satisfies Schwinger-Dyson identity
 in the large-$N$ limit (the loop equation \cite{Migdal:1983gj}):
\eq{\label{mmloop}
W'(\vartheta)=\frac{\lambda}{16\pi^2}\,\int_0^\vartheta 
d\s\,W(\s)W(\vartheta-\s).  } 
The solution 
\cite{Erickson:2000af},\cite{Akemann:2001st},\cite{Zarembo:2001jp} 
of the loop equation is
\eq{
W(\s)=\frac{4\pi}{\sqrt{\lambda}\, \s}
\,I_1\br{\frac{\sqrt{\lambda}\, \s}{2\pi}}.
} 

The integral we need to compute in order to calculate the correlation
function
\rf{ex1dim2} is  the right-hand-side of the
loop equation at $\q=2\pi$. Using properties of the
modified Bessel functions,
\ar{\label{bess}
I'_k(z)&=&\frac{1}{2}\,\br{I_{k-1}(z)+I_{k+1}(z)},
\non
kI_k(z)&=&\frac{z}{2}\,\br{I_{k-1}(z)-I_{k+1}(z)},
} 
we get:
\eq{
\int_0^{\vartheta} d\s\,W(\s)W(\vartheta-\s)
=\frac{32\pi^2}{\lambda\vartheta}\,
I_2\br{\frac{\sqrt{\lambda}\, \vartheta}{2\pi}}.
}
Setting $\q=2\pi$ and substituting into \rf{ex1dim2}, we obtain:
\eq{\label{ex2dim2}
\vev{W(C)\,\o^{ij}}= \frac{1}{N}\,   
2\sqrt{2}\,
I_2\br{\t}\br{\q^i\q^j-\frac{1}{6}\D^{ij}}
\,\frac{R^2}{L^4}\,.
} 
Dividing by the vacuum expectation value of the Wilson loop,
\eq{\label{wvev}
\vev{W(C)}=W(2\pi)=\frac{2}{\t}\,I_1\br{\t},
} 
we arrive at the result \rf{exd2} which we quoted earlier.

\subsection{Chiral primary operators}

The correlator of the Wilson loop with the CPO
of dimension $k$ contains an integral over $k-1$ endpoints of the
scalar propagators (one integration yields an overall factor of $2\pi$):
\ar{
\co&=&\frac{1}{N}\,
\frac{(8\pi^2)^{k/2}}{\sqrt{k}\,\lambda^{k/2}}\,
\frac{\lambda^k}{(8\pi^2L^2)^k}\,
C^I_{i_1\ldots i_k}\q^{i_1}\ldots\q^{i_k}\,
R^k\,
\non &&\times
2\pi\int_0^{2\pi}d\s_1\ldots\int_0^{\s_{k-2}}d\s_{k-1}\,
W(\s_{k-1})W(\s_{k-2}-\s_{k-1})\ldots W(2\pi-\s_1).
}
It is useful to introduce 
\eq{
F_k(\s)=\int_0^{\s}d\s_1\ldots\int_0^{\s_{k-2}}d\s_{k-1}\,
W(\s_{k-1})W(\s_{k-2}-\s_{k-1})\ldots W(\s-\s_1).
}
The correlator is expressed in terms of $F_k(2\pi)$ as 
\eq{\label{corr}
\co=\frac{1}{N}\,
\frac{2\pi}{\sqrt{k}\,(8\pi^2)^{k/2}}\,\lambda^{k/2}F_k(2\pi)
\,\frac{R^k}{L^{2k}}\,Y^I(\q).
}

To find the functions $F_k(\s)$, we again use the loop equation.
 Differentiating $F_k(\s)$ and using \rf{mmloop}, we get 
the recurrence relations:
\eq{\label{rec}
F'_k(\s)=F_{k-1}(\s)+\frac{\lambda}{16\pi^2}\,F_{k+1}(\s),
}
which are
supplemented by initial conditions
\eq{
F_1(\s)=W(\s),~~~F_0(\s)=0.
}
These unambiguously determine all $F_k$. A systematic way to solve these 
recurrence relations is to introduce a generating
function and then use a Laplace transform to convert differential equations 
into algebraic equations. 
The result is

\eq{\label{fk}
F_k(\s)=\frac{k}{\s}\,\br{\frac{4\pi}{\t}}^kI_k\br{\frac{\t\,\s}{2\pi}}.
}
It is straightforward to check that this expression solves
 the recurrence relations 
with the help of \rf{bess}.

Substituting \rf{fk} into \rf{corr}, we obtain
\eq{
\co=\frac{1}{N}\,
2^{k/2}\sqrt{k}\,I_k\br{\t}
\,\frac{R^k}{L^{2k}}\,Y^I(\q).
}
Normalizing by the Wilson loop expectation value \rf{wvev}, we
get \rf{ex}.

\newsection{Radiative Corrections}

\begin{figure}[h]
\hspace*{3cm}
\epsfxsize=10cm
\epsfbox{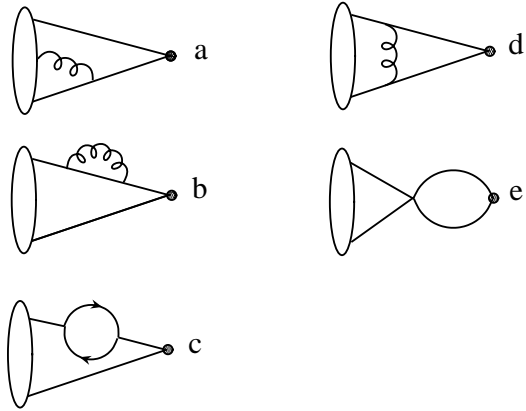}
\caption[x]{\small Leading radiative corrections 
to the correlator of the Wilson loop operator with
the $k=2$ CPO.  These are not taken into account by the sum over
planar rainbow graphs which were computed in Sec.3.}
\label{diagr2}
\end{figure}

The leading radiative corrections come form the Feynman diagrams
which are shown in fig.\ref{diagr2}\footnote{The diagram similar to e, 
but with scalar lines replaced by gluon propagators does not
contribute because of R-charge conservation.}.

Each of these diagrams is separately divergent and regularization is
required to define them properly.  We use a regularization by
dimensional reduction which was previously used in
ref.~\cite{Erickson:2000af}.  The essential observation is that \N SYM
is obtained by dimensional reduction of ten dimensional SYM.  This
dimensional reduction retains sixteen supersymmetries in any
dimension.  Thus, a supersymmetric dimensional regularization of \N
SYM theory is obtained by dimensionally reducing ten dimensions to
$4-\epsilon$ dimensions.  

In this dimensional regularization, the diagram in fig.\ref{diagr2}a
is of higher order than the relevant leading power, $R^2/L^4$, and
therefore does not contribute to ${\cal C}_2$.

\begin{figure}[h]
\hspace*{3cm}
\raisebox{-2cm}[4cm][-2cm]{
\epsfysize=6cm
\epsfbox{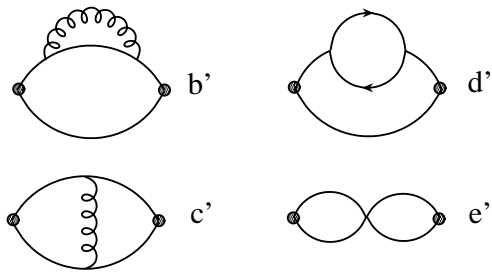}}
\caption[x]{\small The contribution to the leading term in 
$R^2/L^4$ of the diagrams in fig.\ref{diagr2}b-e
are given by these 2-loop diagrams.  This combination of 
diagrams is known to vanish due to  the non-renormalization
theorem of the 2-point function of the CPO.}
\label{diagr3}
\end{figure}

In the limit $L\gg R$, using dimensional regularization, the leading,
$R^2/L^4$, contributions of the remaining diagrams in fig.\ref{diagr2}
can be seen to be identical to results of computing the diagrams which
are displayed in fig.\ref{diagr3}.
This sum of diagrams is known to vanish when the dimension is exactly
four, due to the non-renormalization theorem for the two-point
function of the CPO.  This non-renormalization results from super-conformal
invariance.

Similar arguments apply to the higher CPO's for
which similar non-renormalization theorems can be applied
\cite{D'Hoker:1999tz}.  

This is by no means a proof that all radiative corrections vanish.
But the excellent agreement with strong coupling AdS/CFT results gives
optimism that it is indeed the case.

\newsection{Remarks}

Our main results \rf{exd2} and \rf{ex} are valid when the distance
from operator insertion to the loop is much larger than the loop's
radius.  However, it is not hard to restore the dependence on the
radius and on the orientation of the circle. Consider first the case
when the operator is inserted at the symmetry axis of the circle.  As
follows from the discussion in the beginning of Sec.~\ref{rrd}, we can
find the correlators of the Wilson loop with CPOs at any $R$ and $L$,
not only at large distances, simply by replacing $L^2$ with $L^2+R^2$.
As expected, this is perfectly consistent with the supergravity
prediction, which actually is known in complete generality. If we
denote the displacement of the operator insertion from the axis of
symmetry by $r$ and, as before, $L$ is the distance from the plane of
the circle, the correlation function is obtained from the
large-distance asymptotics by replacing
\eq{\label{disp}
\frac{R^k}{L^{2k}}\,\rightarrow\,
\frac{R^k}{\left[(L^2+r^2-R^2)^2+4L^2R^2\right]^{k/2}}\,.
}
At $r=0$, the denominator on the right hand side indeed
coincides with $(L^2+R^2)^k$. 

We were not able to compute the sum
of rainbow diagrams for $r\neq 0$, but the validity of the prescription
\rf{disp} can be proved by an indirect argument. The point is that, 
once we know the correlators at $r=0$, the dependence on $r$ is
unambiguously fixed by conformal invariance, because
$r$ can be always set to zero by a special conformal transformation,
which maps a circle onto a circle. This transformation is 
 not anomalous and therefore does not affect the dependence 
of the correlator on $\lambda$.

Another remark concerns the dependence of the OPE coefficients on $k$.
It was argued on general grounds that coupling of a Wilson loop to
states of very large spin should be factorially suppressed
\cite{Zarembo:1999bu}.  In a theory with an exponential density of
states, unsuppressed coupling to such states would lead to a
catastrophe in the pair correlator of Wilson loops similar to the
Hagedorn transition. The CPOs we consider in this paper carry the spin
of $SO(6)$ R-symmetry which for $\o^I_k$ is equal to $k$. At weak
coupling, the OPE coefficients are indeed suppressed at large $k$, as
follows from
\rf{wc}, but the supergravity result \rf{sc} seems to
suggest that this suppression disappears at strong coupling.  Careful
inspection of the exact OPE coefficients \rf{ex} shows that limits
$\lambda\rightarrow\infty$ and $k\rightarrow\infty$ do not commute and
coupling of the Wilson loop to operators of very high spin is always
suppressed, but the suppression begins with operators of parametrically
large spin $k\sim \lambda$.  At $k\gg\lambda$:
\eq{
\c_k\propto 2^{-k/2-1}\,\frac{\sqrt{k}}{k!}\,
\frac{\lambda^{(k+1)/2}}{I_1\br{\t}}.
}

\newsection{Discussion}

We have calculated the OPE coefficients of the circular Wilson loops
by re-summation of the planar Feynman graphs without internal vertices.
It is likely that other diagrams  cancel to all orders of perturbation theory.
We have successfully checked this conjecture up to two loops.
Complete agreement of an infinite set of
OPE coefficients with the supergravity
predictions at strong coupling strongly suggests that our 
results  are indeed exact in the 't~Hooft limit.
It would be interesting to see if the arguments
 of Ref.~\cite{Drukker:2000rr} based on diagram-by-diagram
conformal transformations can be invoked 
to prove that only rainbow diagrams contribute to
the OPE coefficients of the circular Wilson loop with CPOs. 

Our result for the OPE coefficients, along with the known exact
expression for the vacuum expectation value of the circular loop, can
be regarded as a prediction for the string theory in $AdS_5\times
S^5$. The usual $\alpha'$ expansion of the world-sheet sigma-model
then coincides with the expansion in $1/\t$:
\eq{\label{expan}
\nco=\frac{1}{N}\,
2^{k/2-1}\sqrt{k\lambda}\,
\br{1-\frac{k^2-1}{2\t}+\frac{k^4-4k^2+3}{8\lambda}+
O\br{\frac{1}{\lambda^{3/2}}}}
\,
\frac{R^k}{L^{2k}}\,Y^I(\q).
} The calculation of the stringy correction to the expectation of the
Wilson loop is a hard problem analogous to instanton calculations in
field theory (see \cite{Drukker:2000ep}, for details).  As usual, such
problems require delicate treatment of various normalization factors
associated with zero modes and with regularization of fluctuation
determinants. However, in the ratio of the expectation values
\rf{expan}, these normalization factors cancel. For this reason, a
calculation of stringy corrections to the OPE coefficients seems less
complicated and perhaps can be accomplished without tremendous
effort.

It would also be interesting to consider similar correlators of Wilson
loops with different contours \cite{Erickson:2000qv} or with other
operators, where some preliminary results are contained in
ref.\cite{Danielsson:1999wt},\cite{Erickson:1999uc}.

\subsection*{Acknowledgments}

We are grateful to Jan Plefka, Matthias Staudacher and Arkady Tseytlin
for useful comments. This work was supported by NSERC of Canada and
NATO Collaborative Linkage Grant SA(PST.CLG.977361)5941.  The work of
K.Z. was also supported in part by the Pacific Institute for the
Mathematical Sciences and in part by RFBR grant 01-01-00549 and RFBR
grant 00-15-96557 for the promotion of scientific schools.

\end{document}